\documentclass[journal]{IEEEtran}

\usepackage{amsmath,amssymb,amsfonts}
\usepackage{graphicx}
\usepackage{booktabs}
\usepackage{cuted}
\usepackage{CJK}
\usepackage{indentfirst}
\usepackage{xcolor}
\usepackage{hyperref}
\usepackage{epstopdf}
\usepackage{bm}
\usepackage{cases}
\usepackage{bookmark}
\usepackage{multirow} % Required for multirows
\usepackage{makecell}

\usepackage{algorithmic}
\usepackage{algorithm}%format of the algorithm
\usepackage{array}
\usepackage{textcomp}
\usepackage{stfloats}
\usepackage{url}
\usepackage{verbatim}
\usepackage{cite}
\hypersetup{
colorlinks=true,
linkcolor=blue,
citecolor=blue
}
\usepackage{placeins}

\usepackage{tikz}
\usetikzlibrary{arrows.meta,positioning,fit,calc}

\newcolumntype{Y}{>{\centering\arraybackslash}X}

\title{HF Single-Station Localization via Multimode Ray Tracing and
Time-Frequency Information Fusion}
\author{
Qinglin~Li,
Wen~Liu,
Yun~Ning,
Zhongxin~Deng,
Junkang~He,
Mingshuo~Liu,
and~Chuangjian~Xia%
\thanks{%
This work has been submitted to the IEEE for possible
publication. Copyright may be transferred without notice,
after which this version may no longer be accessible.

This work was supported in part by the National Key
Research and Development Program of China under
Grant 2023YFA1009100.
(Corresponding author: Wen Liu.)
}%
\thanks{Qinglin Li is with the School of Mathematics and Computational Science, Xiangtan University, Xiangtan 411105, China (Email: 202331510123@smail.xtu.edu.cn).}%
\thanks{Wen Liu, Yun Ning, Zhongxin Deng, Junkang He, Mingshuo Liu, and Chuangjian Xia are with the School of Automation and Electronic Information, Xiangtan University, Xiangtan 411105, China (Email: l\_wen9209@xtu.edu.cn).}%
}

\begin{document}
\maketitle

\begin{abstract}
High-frequency (HF) single-station localization estimates an emitter location from measured
directions of arrival (DOAs) and an ionospheric propagation model. When the
propagation mode composition cannot be identified from conventional
direction-finding (DF) measurements, the same DF observation may produce
multiple physically valid ground landing points under different propagation
modes, making the location inversion ambiguous. This paper proposes a
multimode time-frequency information fusion localization method (MTFIF-L).
Rather than assuming a propagation mode a priori, MTFIF-L retains valid
landing points generated under different propagation modes and uses repeated
observations across time windows and, when available, multiple operating
frequencies to progressively constrain the resulting candidate locations.
Experiments on four measured HF links examine the respective roles of
multimode candidate retention, cross-window processing, and cross-frequency
processing. All four links yield unique location decisions. Links A--C
achieve relative errors of $1.40\%$, $0.78\%$, and $3.41\%$, respectively.
Link D, the only single-frequency case, also yields a unique decision from
repeated temporal observations, although its post-localization error is
substantially larger. Propagation-path analysis further reveals link-dependent
differences in reflection layer, hop count, magnetoionic mode, and elevation
angle among the ray-tracing paths associated with the measured links.
Overall, the results show that repeated temporal observations can support
localization within an individual frequency, while multifrequency information
provides additional robustness against unreliable frequency-specific
outcomes, enabling HF single-station localization under unknown and
potentially coexisting propagation modes while retaining interpretable
propagation information for post-localization analysis.
\end{abstract}


\section{Introduction}
\label{sec:introduction}

High-frequency (HF) radio can exploit ionospheric propagation to achieve 
long-range beyond-line-of-sight communication 
and is widely used in communications, spectrum monitoring, and electronic reconnaissance
\cite{Davies1990,Ratcliffe1959,Hargreaves1992}.
Accurate localization of HF emitters provides 
an important basis for interference identification, 
monitoring of unauthorized transmissions, and situational awareness
\cite{Heitmann2013}.
Conventional localization techniques generally rely on multiple geographically distributed receiving stations, imposing stringent requirements on interstation baselines, time synchronization, and multistation coordination.
When deployment is constrained or only a single receiving station can reliably receive the target signal, HF single-station localization (SSL) remains of considerable practical importance
\cite{Erhel2007}.

HF SSL typically estimates the emitter location by combining the direction of arrival (DOA) measured at the receiving station with an ionospheric propagation model.
The DOA provides the azimuth and elevation angles at which the radio wave arrives at the receiver.
Angular measurements alone, however, cannot determine the ground location of a distant emitter.
A propagation model is additionally required to describe how the radio wave reaches the receiver through the ionosphere and thereby establish the relationship between the measured DOA and the corresponding ground location.
Consequently, HF single-station location inversion depends 
not only on DOA measurement accuracy 
but also on the ionospheric propagation paths associated with the received signal.

The propagation path of an HF skywave is generally not unique.
The layered electron density structure of the ionosphere, anisotropy induced by the geomagnetic field, and multihop propagation between the ground and the ionosphere allow radio waves to propagate along multiple physically feasible paths involving different reflection layers, ordinary/extraordinary (O/X) magnetoionic modes, and hop counts
\cite{Davies1990,Ratcliffe1959,Hargreaves1992,Davey2019TAES}.
Under the same DF observation condition, different propagation modes may correspond to different ground landing points.
Conventional HF direction-finding data generally do not identify the propagation mode composition associated with the received signal.
Consequently, a single DF observation may correspond to multiple physically valid candidate locations, preventing the emitter location from being uniquely determined by a single location inversion.

\subsection{Related Work}

DOA accuracy and propagation modeling are the two basic determinants of HF
single-station location inversion. HF direction-finding measurements can be
affected by array configuration, site environment, antenna pattern errors,
channel mismatch, noise, interference, multipath superposition, and
ionospheric disturbances \cite{Milatos1993,Haines1992,Lai2017}.
Polarization-sensitive arrays, improved angle estimators, phase ambiguity
resolution, and antenna pattern calibration have therefore been used to
improve azimuth/elevation estimates
\cite{Erhel2007,Tsai2014,Lai2017}. On the propagation side, early methods
relied mainly on equivalent virtual height relations
\cite{Breit1926,Martyn1935}, whereas later approaches incorporated
ionospheric models and numerical ray tracing to map measured DOAs and
operating frequencies to ground locations
\cite{Xiong2023TAES,Lai2008,Huang2006,Tsai2014,Zhu2016,Li2026TAP}.
Ionospheric variability has also been treated through joint estimation of
propagation and target parameters in skywave radar \cite{Ahmed2022TAES}.
These studies improve the angular observations or the propagation-path
calculation, while the propagation mode used for ray-tracing inversion is
commonly prescribed or treated as known.

A complementary line of work exploits separable multipath components.
Fabrizio and Heitmann combined cone angle and intercomponent delay
information to constrain independently obtained locations
\cite{Heitmann2013}. Such approaches generally require reliable component
separation using spatially separated antennas, super-resolution or blind
processing, or intercomponent delay extraction
\cite{Johnson1994,Heitmann2013}; closely spaced HF multipath components can
nevertheless remain difficult to separate \cite{Tang2022}. They are
therefore most applicable when component-level propagation parameters are
directly available.

For conventional HF direction-finding data containing only azimuth,
elevation, and operating frequency, the propagation mode composition is
generally unavailable before inversion. Ray-tracing-based localization
methods that prescribe a propagation mode therefore make the inferred
ground location dependent on the adopted propagation mode. This leaves HF
single-station localization under unknown and potentially coexisting
propagation modes insufficiently addressed.

\subsection{Motivation and Contributions}

For a stationary emitter, its geographic location remains unchanged across
observations, whereas the ionospheric background and the contributing
propagation paths may vary with time and operating frequency. This provides
a physical basis for retaining the propagation mode, time window, and
frequency associations of multimode ray-tracing candidates and using these
associations to determine the common emitter location.

Accordingly, we propose MTFIF-L for HF single-station localization without
requiring the propagation mode composition to be identified a priori.
We preserve the propagation mode, time window, and operating frequency 
associated with each physically valid ray-tracing candidate, 
and use these associations to determine the common emitter location.

Our main contributions are summarized as follows:
\begin{enumerate}

\item We constructed a physics-guided multimode candidate representation
through multimode ionospheric ray tracing, preserving the propagation mode,
time window, and operating frequency associated with each physically valid
ground landing.

\item We developed MTFIF-L to estimate the emitter location from multimode
candidate sets by exploiting temporal and frequency information under both
single-frequency and multifrequency observation conditions.

\item We evaluated MTFIF-L on four measured HF links with different
frequency and propagation conditions through mechanism analysis,
localization accuracy evaluation, post-processing runtime measurement,
and propagation-path interpretation.

\end{enumerate}

\section{Methodology}

In this section, we formulate the HF single-station localization problem,
construct the physics-guided multimode candidate representation, and
develop the proposed localization procedure using time and frequency
information. We summarize the complete processing flow at the end of the
section.

\subsection{HF Single-Station Localization Model}
\label{subsec:hf_ssl_model}

Consider a fixed HF receiving station at
\begin{equation}
\mathbf q_r=(\phi_r,\lambda_r,h_r),
\end{equation}
and a stationary ground transmitter at the unknown location
\begin{equation}
\mathbf q_s=(\phi_s,\lambda_s,h_s),
\end{equation}
where $\phi$, $\lambda$, and $h$ denote latitude, longitude, and altitude,
respectively. Because the transmitter is ground-based, its altitude is treated
as a fixed site parameter rather than an independent search variable.
For the $i$th direction-finding (DF) observation,
\begin{equation}
\mathbf o_i
=
(t_i,f_i,\alpha_i,\beta_i),
\qquad
i=1,\ldots,N,
\label{eq:df_observation}
\end{equation}
where $t_i$ and $f_i$ denote the observation time and operating frequency,
and $\alpha_i$ and $\beta_i$ denote the measured azimuth and elevation
angles. The localization objective is to estimate the fixed transmitter
location $\mathbf q_s$ from the DF observations and a physical HF
propagation model.

Conventional DF measurements do not identify which propagation modes
contribute to a received signal. Let $\mathcal M$ denote the propagation
modes considered in the inversion. For observation $\mathbf o_i$, one or
more modes may physically coexist; their actual but unavailable set is
denoted by
\begin{equation}
\varnothing
\neq
\mathcal M_i^{\mathrm{act}}
\subseteq
\mathcal M,
\label{eq:active_mode_set}
\end{equation}
with no restriction that
$|\mathcal M_i^{\mathrm{act}}|=1$.
For each considered mode $m\in\mathcal M$, ionospheric ray tracing
associates the measured arrival condition with a mode-dependent ground
landing $\mathbf p_i^m$ under the propagation background adopted for
observation time $t_i$. Only physically valid ground landings are
retained.
Accordingly, the spatial inversion of one DF observation is represented by
the candidate space set
\begin{equation}
\mathcal P_i
=
\left\{
\left(
\mathbf p_i^m,
m,
t_i,
f_i
\right)
\,\middle|\,
m\in\mathcal M,\;
\mathbf p_i^m\ \text{is valid}
\right\},
\label{eq:observation_candidate_space}
\end{equation}
rather than by a unique geographic point.

If the physical associations in $\mathcal P_i$ are discarded and all
mode-dependent landing points are pooled before localization, the resulting
spatial multiset is
\begin{equation}
\mathcal P_{\mathrm{pool}}
=
\biguplus_{i=1}^{N}
\biguplus_{\substack{m\in\mathcal M\\
\mathbf p_i^m\ \mathrm{valid}}}
\left\{
\mathbf p_i^m
\right\},
\label{eq:pooled_candidate_points}
\end{equation}
where $\biguplus$ denotes multiset union. For an arbitrary geographic
region $\mathcal A$, define
\begin{equation}
\begin{aligned}
n_i(\mathcal A)
&=
\sum_{\substack{m\in\mathcal M\\
\mathbf p_i^m\ \mathrm{valid}}}
I\!\left[
\mathbf p_i^m\in\mathcal A
\right],\\
C_{\mathrm{cand}}(\mathcal A)
&=
\sum_{i=1}^{N}
n_i(\mathcal A),\\
C_{\mathrm{obs}}(\mathcal A)
&=
\sum_{i=1}^{N}
I\!\left[
n_i(\mathcal A)>0
\right],
\end{aligned}
\label{eq:candidate_observation_counts}
\end{equation}
where $I[\cdot]$ is the indicator function, $C_{\mathrm{cand}}(\mathcal A)$ counts the mode-dependent candidate
landings in $\mathcal A$, whereas $C_{\mathrm{obs}}(\mathcal A)$ counts the
DF observations that contribute at least one candidate to that region.
They satisfy
\begin{equation}
C_{\mathrm{obs}}(\mathcal A)
\le
C_{\mathrm{cand}}(\mathcal A)
\le
|\mathcal M|\,
C_{\mathrm{obs}}(\mathcal A).
\label{eq:candidate_observation_count_relation}
\end{equation}
Thus, $C_{\mathrm{cand}}(\mathcal A)$ can exceed the number of supporting
DF observations because a single observation may contribute multiple
mode-dependent landings to the same region.

Moreover, direct spatial pooling applies the projection
\begin{equation}
\left(
\mathbf p_i^m,
m,
t_i,
f_i
\right)
\overset{\Pi_{\mathrm{sp}}}{\longmapsto}
\mathbf p_i^m.
\label{eq:spatial_projection}
\end{equation}
The projection preserves the landing location but discards its propagation
mode, observation time, and operating frequency associations. Consequently,
clustering $\mathcal P_{\mathrm{pool}}$ uses only spatial proximity and
cannot exploit these physical associations.

Accordingly, we formulate the localization problem using
$\{\mathcal P_i\}_{i=1}^{N}$ with their physical associations retained,
rather than using the pooled landing points. We seek the common stationary
transmitter location from these candidate sets.

Based on this formulation, we next organize the DF observations into time
windows, generate the multimode candidate sets, and exploit their temporal
and frequency associations for localization.

\subsection{Physics-Guided Candidate Representation}
\label{subsec:physics_guided_candidate}

We construct the localization candidates in two steps. We first preprocess
the DF observations to obtain distinct measurement states while preserving
their observation multiplicities. We then apply multimode ionospheric ray
tracing to these states and retain the physically valid ground landings to
form $\mathcal C_{f,k}$.

\subsubsection{DF Observation Preprocessing}
\label{subsec:df_preprocessing}

Let $\mathcal O=\{\mathbf o_i\}_{i=1}^{N}$ denote the complete set of
raw DF observations, with $\mathbf o_i$ defined in
\eqref{eq:df_observation}. The observation period is partitioned into
nonoverlapping time windows of duration $\Delta T=5~\mathrm{min}$:
\begin{equation}
\begin{aligned}
I_k
&=
[t_0+(k-1)\Delta T,\,t_0+k\Delta T),\\
\mathcal W_k
&=
\left\{
\mathbf o_i\in\mathcal O
\,\middle|\,
t_i\in I_k
\right\},
\qquad k=1,\ldots,K.
\end{aligned}
\label{eq:time_window_set}
\end{equation}
The propagation background adopted for ray tracing in the $k$th window is
denoted by $\mathcal I_k$ and is held fixed for all observations with
$t_i\in I_k$.

Within each time window, the measured azimuths are screened before ray
tracing. Because azimuth is circular, each $\alpha_i$ is mapped to the
unit circle as
\begin{equation}
\mathbf u_i
=
\begin{bmatrix}
\cos\alpha_i\\
\sin\alpha_i
\end{bmatrix},
\qquad
\mathbf o_i\in\mathcal W_k.
\label{eq:azimuth_embedding}
\end{equation}
Density-based spatial clustering of applications with noise (DBSCAN) is then
applied in this unit-circle embedding using the Euclidean
chord distance threshold $\varepsilon_{\rm DB}=2\sin(5^\circ/2)$ and
$\mathrm{MinPts}_k=\max(5,\lceil0.05|\mathcal W_k|\rceil)$, yielding
clusters $\{\mathcal G_{k,j}\}_{j=1}^{M_k}$ ranked by decreasing size.
The largest cluster is retained as the main directional cluster, and,
when at least two clusters are present, the second-largest cluster is the
only secondary cluster candidate. Let $N_1$ and $N_2$ denote their sizes.
The secondary cluster is retained only if
\begin{equation}
\begin{aligned}
N_2
&\geq
\max\left(20,\operatorname{round}(0.15N_1)\right),\\
\operatorname{MAD}_{\rm circ,2}
&\leq 2^\circ,
\qquad
\Delta\alpha_{12}\leq 35^\circ.
\end{aligned}
\label{eq:secondary_cluster_retention}
\end{equation}
where $\operatorname{MAD}_{\rm circ,2}$ denotes the circular median absolute
deviation (MAD) of the secondary-cluster azimuths and $\Delta\alpha_{12}$ is the circular
separation between the circular-median azimuths of the two clusters.
Let $\mathcal J_k$ denote the retained cluster indices. If the window
contains too few observations for DBSCAN or no valid cluster is obtained,
all observations in that window are retained. The observations passing the
directional screening are
\begin{equation}
\widetilde{\mathcal W}_k
=
\left\{
\mathbf o_i\in\mathcal W_k
\,\middle|\,
\mathbf u_i\in
\bigcup_{j\in\mathcal J_k}\mathcal G_{k,j}
\right\}.
\label{eq:retained_window_set}
\end{equation}

Because observations with identical $(f,\alpha,\beta)$ in the same window
produce identical ray-tracing inputs, they are represented by the distinct
DF states
\begin{equation}
\begin{aligned}
\left\{
\boldsymbol{\xi}_{u,k}
\right\}_{u=1}^{U_k}
&=
\operatorname{unique}
\left\{
(f_i,\alpha_i,\beta_i):
\right.\\[-1mm]
&\hspace{16mm}
\left.
\mathbf o_i\in\widetilde{\mathcal W}_k
\right\},\\
\boldsymbol{\xi}_{u,k}
&=
(f_{u,k},\alpha_{u,k},\beta_{u,k}).
\end{aligned}
\label{eq:distinct_df_states}
\end{equation}
For each distinct state, its observation multiplicity is
\begin{equation}
N_{u,k}^{\mathrm{obs}}
=
\sum_{\mathbf o_i\in\widetilde{\mathcal W}_k}
I\!\left[
(f_i,\alpha_i,\beta_i)
=
\boldsymbol{\xi}_{u,k}
\right],
\label{eq:observation_multiplicity}
\end{equation}
and the preprocessed observations in the $k$th time window are represented as
\begin{equation}
\widehat{\mathcal O}_k
=
\left\{
\left(
\boldsymbol{\xi}_{u,k},
N_{u,k}^{\mathrm{obs}}
\right)
\right\}_{u=1}^{U_k}.
\label{eq:preprocessed_df_set}
\end{equation}
Thus, only one ray-tracing calculation is required for each distinct DF
state, while its raw observation multiplicity is retained for the subsequent
window-level localization objective.

\subsubsection{Candidate Generation Based on Multimode Ray Tracing}
\label{subsubsec:multimode_candidate_generation}
We convert the distinct DF states retained after preprocessing into
mode-dependent ground landing candidates through multimode ionospheric
ray tracing. We consider the propagation mode set
\begin{equation}
\mathcal M
=
\{O_1,X_1,O_2,X_2\},
\label{eq:propagation_mode_set}
\end{equation}
where $O$ and $X$ denote the ordinary and extraordinary magnetoionic
waves, respectively, and the subscripts denote the propagation hop
number. Propagation beyond two hops is not considered in this work.

For each distinct DF state $\boldsymbol{\xi}_{u,k}$ in the $k$th time
window, we trace one O mode and one X mode under the
corresponding propagation background $\mathcal I_k$:
\begin{equation}
\mathcal R_{u,k}^{P}
=
\operatorname{RT}
\left(
\mathbf q_r,
\boldsymbol{\xi}_{u,k},
P;
\mathcal I_k
\right),
\qquad
P\in\{O,X\},
\label{eq:multimode_ray_tracing}
\end{equation}
where $\operatorname{RT}(\cdot)$ denotes the numerical ray-tracing
operation and $\mathcal R_{u,k}^{P}$ is the resulting ray path. The ray is
launched from $\mathbf q_r$ using the operating frequency and measured
arrival angles contained in $\boldsymbol{\xi}_{u,k}$, while $\mathcal I_k$
specifies the propagation background for the $k$th time window.

From each physical O/X-mode ray, we extract the first and second
physically valid ground landings as
\begin{equation}
\mathbf p_{u,k}^{P_h}
=
\operatorname{Land}_h
\left(
\mathcal R_{u,k}^{P}
\right),
\qquad
P\in\{O,X\},\quad
h\in\{1,2\},
\label{eq:ray_landing_extraction}
\end{equation}
where $\operatorname{Land}_h(\cdot)$ returns the $h$th valid ground
landing of the traced ray and $P_h\in\mathcal M$ denotes the associated
propagation mode. Thus, the one- and two-hop candidates for each
magnetoionic wave are obtained from a single two-hop physical ray trace.
Only physically valid ground landings are retained.

For each frequency $f$ and time window $k$, we collect all valid landing
candidates generated from the DF states at that frequency as
\begin{equation}
\begin{aligned}
\mathcal C_{f,k}
=\Bigl\{
(\mathbf p_{u,k}^{m},m,N_{u,k}^{\mathrm{obs}})
\ \Big|\ &
f_{u,k}=f,\;m\in\mathcal M,\\
&\mathbf p_{u,k}^{m}\ \text{is valid}
\Bigr\}.
\end{aligned}
\label{eq:window_candidate_set}
\end{equation}
Hence, each element of $\mathcal C_{f,k}$ retains three pieces of
information required by the subsequent localization procedure: the
ground landing location, its propagation mode, and the observation
multiplicity associated with the corresponding DF state.

For notational convenience in the subsequent localization formulation,
the candidates in $\mathcal C_{f,k}$ are indexed as
\begin{equation}
\mathcal C_{f,k}
=
\left\{
c_{\ell}^{f,k}
\right\}_{\ell=1}^{L_{f,k}},
\qquad
L_{f,k}
=
\left|
\mathcal C_{f,k}
\right|,
\label{eq:indexed_candidate_set}
\end{equation}
where
\begin{equation}
c_{\ell}^{f,k}
=
\left(
\mathbf p_{\ell}^{f,k},
m_{\ell}^{f,k},
N_{\ell}^{\mathrm{obs}}
\right).
\label{eq:indexed_candidate}
\end{equation}
Here, $\ell$ is only an index for the candidates in
$\mathcal C_{f,k}$. The resulting candidate sets $\mathcal C_{f,k}$ constitute
the inputs to the localization procedure described in
Section~\ref{subsec:mtfif_localization}.

\subsection{Multimode Localization Using Time and Frequency Information Fusion}
\label{subsec:mtfif_localization}

Given the candidate sets $\mathcal C_{f,k}$ constructed above, we develop
the localization procedure that maps the multimode ray-tracing candidates
to the final set-valued output through
\begin{equation}
\mathcal C_{f,k}
\longrightarrow
\left\{
\mathcal H_{f,k,m},
\mathcal H_{f,k}^{U}
\right\}
\longrightarrow
\mathcal S_f
\longrightarrow
\widehat{\mathcal Q}_s.
\end{equation}
We first extract window-level regional minimum location sets for each
frequency, time window, and propagation mode. We then evaluate these sets
across time windows at the same frequency to form the frequency-specific
competing location set $\mathcal S_f$. Finally, we apply the
frequency-dependent decision to obtain $\widehat{\mathcal Q}_s$.

\subsubsection{Window-Level Mode-Specific Location Extraction}
For each frequency $f$ and time window $k$, we first separate the
candidates in $\mathcal C_{f,k}$ by propagation mode to preserve their
distinct physical interpretations. For each $m\in\mathcal M$, define
\begin{equation}
\mathcal C_{f,k,m}
=
\left\{
c_{\ell}^{f,k}\in\mathcal C_{f,k}
\,\middle|\,
m_{\ell}^{f,k}=m
\right\}.
\label{eq:mode_specific_candidate_subset}
\end{equation}

Because the transmitter altitude $h_s$ is fixed, the localization
search is performed only over the horizontal geographic coordinates.
Accordingly, a receiver-centered polar coordinate system represents a
search location as $\mathbf q(r,\theta)$, where $r$ and $\theta$
denote the ground range and bearing relative to the receiver,
respectively, and $\mathbf q=(\phi_q,\lambda_q)$ denotes the
horizontal geographic component of a candidate transmitter location.
For a candidate landing location
$\mathbf p_{\ell}^{f,k}=(\phi_{\ell},\lambda_{\ell})$, their
great-circle distance is
\begin{equation}
\begin{aligned}
d_{\ell}^{f,k,m}(\mathbf q)
=R_e\arccos\Bigl[&
\sin\phi_q\sin\phi_{\ell}\\
&+\cos\phi_q\cos\phi_{\ell}
\cos(\lambda_q-\lambda_{\ell})
\Bigr],
\end{aligned}
\label{eq:mode_specific_candidate_distance}
\end{equation}
where $R_e$ denotes the Earth radius. The same great-circle metric between
arbitrary geographic locations is denoted by $d_{\mathrm g}(\cdot,\cdot)$
hereafter.

To preserve observation multiplicity, we replicate each candidate distance
according to $N_{\ell}^{\mathrm{obs}}$, forming the distance multiset
\begin{equation}
\mathcal D_{f,k,m}(\mathbf q)
=
\biguplus_{c_{\ell}^{f,k}\in\mathcal C_{f,k,m}}
\underbrace{
\left\{
d_{\ell}^{f,k,m}(\mathbf q),
\ldots,
 d_{\ell}^{f,k,m}(\mathbf q)
\right\}
}_{N_{\ell}^{\mathrm{obs}}}.
\label{eq:mode_specific_distance_multiset}
\end{equation}
Here, $\biguplus$ denotes multiset union. We then define the mode-specific
median-distance objective as
\begin{equation}
J_{f,k,m}(\mathbf q)
=
\operatorname{median}
\left(
\mathcal D_{f,k,m}(\mathbf q)
\right).
\label{eq:mode_specific_median_objective}
\end{equation}
The median limits the influence of spatially deviating candidates. Because
$J_{f,k,m}(\mathbf q)$ may contain multiple spatially separated low-value
regions, we retain all numerically distinct regional minima. Here,
$\operatorname{RMin}[J]$ denotes the finite set of refined regional minimum
locations obtained by the coarse-to-fine spatial search over the
stage-specific receiver-centered search domain; the numerical procedure is
detailed in the Appendix. Accordingly,
\begin{equation}
\mathcal H_{f,k,m}
=
\operatorname{RMin}
\left[
J_{f,k,m}(\mathbf q)
\right].
\label{eq:mode_specific_window_locations}
\end{equation}

Because the actual propagation mode remains unavailable at this stage,
we retain all mode-specific location sets through
\begin{equation}
\mathcal H_{f,k}^{U}
=
\bigcup_{m\in\mathcal M}
\mathcal H_{f,k,m}.
\end{equation}
Here, the superscript $U$ denotes the union over all considered
propagation modes. Observation multiplicity is used in constructing
the window-level spatial objectives through
$\mathcal D_{f,k,m}(\mathbf q)$. After the window-level location sets
are obtained, each valid time window contributes one term to the
same-frequency processing, while each valid operating frequency
contributes one comparison unit to the final frequency-dependent
decision.

\subsubsection{Same-Frequency Cross-Window Median-Distance Localization}
At a fixed operating frequency $f$, the transmitter location is invariant
across time windows, whereas the propagation background and the actual
propagation mode may vary. Hence, transmitter-compatible locations should
recur spatially across windows.

Accordingly, we introduce complementary mode-unrestricted and
mode-preserving cross-window formulations, allowing spatial recurrence
with either varying or unchanged propagation mode.

Let $\mathcal K_f$ denote the set of valid time windows at frequency $f$,
\begin{equation}
\mathcal K_f
=
\left\{
k:
\mathcal H_{f,k}^{U}\neq\varnothing
\right\}.
\label{eq:valid_frequency_windows}
\end{equation}
When $|\mathcal K_f|\ge2$, each time window contributes the
distance from a search location $\mathbf q$ to the nearest location in its
multimode set $\mathcal H_{f,k}^{U}$. The corresponding mode-unrestricted
cross-window objective is defined as
\begin{equation}
J_f^{U}(\mathbf q)
=
\operatorname{median}_{k\in\mathcal K_f}
\left[
\min_{\mathbf h\in\mathcal H_{f,k}^{U}}
d_{\mathrm g}(\mathbf q,\mathbf h)
\right].
\end{equation}
Its regional minima are retained as
\begin{equation}
\mathcal B_f^{U}
=
\operatorname{RMin}
\left[
J_f^{U}(\mathbf q)
\right].
\end{equation}
This mode-unrestricted objective allows the nearest propagation mode to vary
across time windows. To additionally retain locations that recur under the
same propagation mode across multiple windows, define
\begin{equation}
\mathcal K_{f,m}
=
\left\{
k:
\mathcal H_{f,k,m}\neq\varnothing
\right\}.
\end{equation}
For each $m\in\mathcal M$ with $|\mathcal K_{f,m}|\ge2$, the corresponding
mode-preserving cross-window objective is
\begin{equation}
J_{f,m}(\mathbf q)
=
\operatorname{median}_{k\in\mathcal K_{f,m}}
\left[
\min_{\mathbf h\in\mathcal H_{f,k,m}}
d_{\mathrm g}(\mathbf q,\mathbf h)
\right],
\end{equation}
and its regional minima are
\begin{equation}
\mathcal B_{f,m}
=
\operatorname{RMin}
\left[
J_{f,m}(\mathbf q)
\right].
\end{equation}
Here, the propagation mode $m$ is kept unchanged across all participating
time windows.

We combine the regional minimum sets obtained from the mode-unrestricted
objective and all valid mode-preserving objectives to form the
frequency-specific competing location set. 
For $|\mathcal K_f|\ge2$,
\begin{equation}
\mathcal S_f
=
\mathcal B_f^{U}
\cup
\bigcup_{\substack{
m\in\mathcal M\\
|\mathcal K_{f,m}|\ge2
}}
\mathcal B_{f,m}.
\end{equation}
If only one valid time window $k_1$ is available, no cross-window
comparison can be performed, and the multimode location set of that window
is retained directly:
\begin{equation}
\mathcal S_f
=
\mathcal H_{f,k_1}^{U}.
\end{equation}
Thus, $\mathcal S_f$ contains the competing locations retained 
after evaluating the window-level location sets available 
at frequency $f$ and serves as the input to the final frequency-dependent decision.

\subsubsection{Frequency-Dependent Localization Decision}

After same-frequency cross-window processing, we use the number of
nonempty frequency-specific competing sets to select the final decision
rule. We define
\begin{equation}
\mathcal F_{\mathrm{loc}}
=
\left\{
f:
\mathcal S_f\neq\varnothing
\right\},
\qquad
N_{\mathrm{loc}}
=
\left|
\mathcal F_{\mathrm{loc}}
\right|.
\end{equation}
If $N_{\mathrm{loc}}=0$, the localization output is
$\widehat{\mathcal Q}_s=\varnothing$. For $N_{\mathrm{loc}}\ge1$, the
corresponding frequency count branch returns either a unique location
decision or the unresolved candidate set.

A singleton is reported as a unique decision only when the corresponding
branch-specific uniqueness rule is satisfied; uniqueness is not inferred
from the cardinality of $\widehat{\mathcal Q}_s$ alone.

For the comparisons used in the three frequency count branches, consider a
finite set of competing locations
\begin{equation}
\mathcal X
=
\left\{
\mathbf x_a
\right\}_{a=1}^{A}
\end{equation}
and a set of comparison units $\mathcal E$. Let $\rho_{e,a}$ denote the
distance associated with location $\mathbf x_a$ and comparison unit
$e\in\mathcal E$. For two competing locations $\mathbf x_a$ and
$\mathbf x_b$, define
\begin{equation}
V_{ab}
=
\sum_{e\in\mathcal E}
I
\left(
\rho_{e,a}
<
\rho_{e,b}
\right),
\end{equation}
where $I(\cdot)$ is the indicator function. An exact tie contributes
to neither side of the comparison. The branch-specific numerical
handling of floating point equivalence is given in
the Appendix. When $A>1$, location
$\mathbf x_{a^\star}$ is the unique strict pairwise winner if
\begin{equation}
V_{a^\star b}
>
V_{b a^\star},
\qquad
\forall b\neq a^\star.
\end{equation}
For $A=1$, the sole member of $\mathcal X$ is unique without further
pairwise comparison. For $A>1$, we distinguish three decision branches
according to $N_{\mathrm{loc}}$.

For $N_{\mathrm{loc}}=1$, we use only the available time window evidence.
For $N_{\mathrm{loc}}=2$, we use reciprocal correspondence and agreement
between the two frequencies because a two-term median distance objective
may be nonunique. For $N_{\mathrm{loc}}\ge3$, we use cross-frequency
median distance consolidation followed by frequency-wise pairwise
comparison.

\paragraph{Single-Frequency Cross-Window Pairwise Decision}
When $N_{\mathrm{loc}}=1$, let $f$ denote the only frequency in
$\mathcal F_{\mathrm{loc}}$, and write
$\mathcal S_f=\{\mathbf s_{f,a}\}_{a=1}^{A_f}$. If fewer than two valid
time windows are available, no cross-window comparison can be established,
and $\mathcal S_f$ is retained as the candidate set.

For $|\mathcal K_f|\ge2$, each valid time window acts as one comparison
unit. The distance from competing location $\mathbf s_{f,a}$ to the
multimode location set of window $k$ is
\begin{equation}
\delta_{f,k,a}
=
\min_{\mathbf h\in\mathcal H_{f,k}^{U}}
d_{\mathrm g}
\left(
\mathbf s_{f,a},
\mathbf h
\right).
\end{equation}
Accordingly, the common pairwise rule is applied with
\begin{equation}
\mathcal X=\mathcal S_f,
\qquad
\mathcal E=\mathcal K_f,
\qquad
\rho_{k,a}=\delta_{f,k,a}.
\end{equation}
Thus, each time window contributes at most one comparison for a given
location pair, while the propagation mode providing the nearest
window-level location may vary across windows. If a unique strict
pairwise winner exists over all valid time windows, that location is
returned as the unique single-frequency location decision. Otherwise,
$\mathcal S_f$ is retained as the candidate set.

\paragraph{Two-Frequency Reciprocal Correspondence Decision}
When $N_{\mathrm{loc}}=2$, let the two frequencies be $f_1$ and $f_2$,
with competing location sets $\mathcal S_{f_1}$ and $\mathcal S_{f_2}$.
Locations that are mutually nearest across the two frequency-specific sets
are paired as
\begin{equation}
\mathcal P_{12}
=
\left\{
\left(
\mathbf s_{f_1,i},
\mathbf s_{f_2,j}
\right)
\,\middle|\,
\begin{array}{l}
\mathbf s_{f_2,j}
\in
\displaystyle
\arg\min_{\mathbf u\in\mathcal S_{f_2}}
d_{\mathrm g}
\left(
\mathbf s_{f_1,i},
\mathbf u
\right),
\\[2mm]
\mathbf s_{f_1,i}
\in
\displaystyle
\arg\min_{\mathbf v\in\mathcal S_{f_1}}
d_{\mathrm g}
\left(
\mathbf v,
\mathbf s_{f_2,j}
\right)
\end{array}
\right\}.
\end{equation}
Each reciprocal pair is represented by its spherical midpoint,
\begin{equation}
\mathbf z_a
=
\operatorname{MidPoint}
\left(
\mathbf s_{f_1,i_a},
\mathbf s_{f_2,j_a}
\right),
\qquad
a=1,\ldots,|\mathcal P_{12}|.
\end{equation}
If only one reciprocal pair exists, its midpoint is returned directly as
the unique two-frequency location decision.

When multiple reciprocal pairs remain, the midpoint of each pair is
evaluated independently at the two frequencies. For
$f\in\{f_1,f_2\}$ and $k\in\mathcal K_f$, define
\begin{equation}
D_{f,k,a}
=
\min_{\mathbf h\in\mathcal H_{f,k}^{U}}
d_{\mathrm g}
\left(
\mathbf z_a,
\mathbf h
\right).
\end{equation}
For each frequency $f$, the common pairwise rule is applied to the midpoint
set with
\begin{equation}
\mathcal X
=
\left\{
\mathbf z_a
\right\}_{a=1}^{|\mathcal P_{12}|},
\qquad
\mathcal E=\mathcal K_f,
\qquad
\rho_{k,a}=D_{f,k,a}.
\end{equation}
A unique two-frequency location decision is returned only when the two
frequencies independently select the same unique reciprocal pair,
\begin{equation}
a_{f_1}^{\star}
=
a_{f_2}^{\star}
=
a^\star.
\end{equation}
The corresponding midpoint $\mathbf z_{a^\star}$ is then reported as the
unique location decision. If the two frequencies do not identify the same
unique pair, the reciprocal pair midpoints are retained as the candidate
set.

\paragraph{Multifrequency Median-Distance and Pairwise Decision}
When $N_{\mathrm{loc}}\ge3$, each frequency first contributes the distance
from a search location $\mathbf q$ to the nearest member of its
frequency-specific competing location set,
\begin{equation}
d_f(\mathbf q)
=
\min_{\mathbf s\in\mathcal S_f}
d_{\mathrm g}
\left(
\mathbf q,
\mathbf s
\right).
\end{equation}
The cross-frequency median distance objective is
\begin{equation}
J_{\mathrm{CF}}(\mathbf q)
=
\operatorname{median}_{f\in\mathcal F_{\mathrm{loc}}}
d_f(\mathbf q),
\end{equation}
and its regional minima form the finite cross-frequency
competing location set
\begin{equation}
\mathcal C_{\mathrm{CF}}
=
\operatorname{RMin}
\left[
J_{\mathrm{CF}}(\mathbf q)
\right].
\end{equation}
The objective $J_{\mathrm{CF}}(\mathbf q)$ is used only to generate
$\mathcal C_{\mathrm{CF}}$ and is not reused to rank its members. If
$\mathcal C_{\mathrm{CF}}$ contains only one location, that location is
returned directly as the unique location decision.

When multiple cross-frequency locations remain, write
$\mathcal C_{\mathrm{CF}}=\{\mathbf c_a\}_{a=1}^{A_{\mathrm{CF}}}$.
Each operating frequency then acts as one comparison unit through
\begin{equation}
\mathcal X=\mathcal C_{\mathrm{CF}},
\qquad
\mathcal E=\mathcal F_{\mathrm{loc}},
\qquad
\rho_{f,a}=d_f(\mathbf c_a).
\end{equation}
Each frequency contributes one ordinal comparison for each candidate pair.
If a unique strict pairwise winner $\mathbf c_{a^\star}$ exists, it is
returned; otherwise, $\mathcal C_{\mathrm{CF}}$ is retained.

Across all three branches, a unique location decision is denoted by
$\hat{\mathbf q}_s$, with
$\widehat{\mathcal Q}_s=\{\hat{\mathbf q}_s\}$.

\subsection{Overall Procedure}
\label{subsec:overall_procedure}

\begin{figure}[htbp]
\centering
\includegraphics[width=0.78\linewidth]
{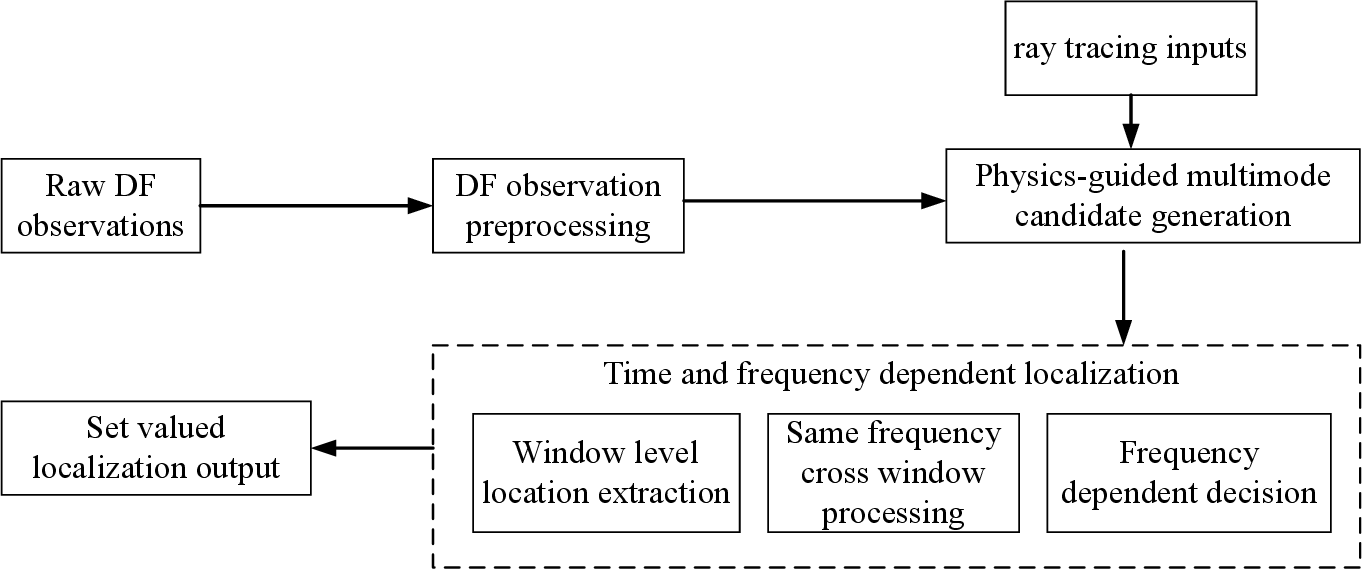}
\caption{Overall processing framework of MTFIF-L.}
\label{fig:overall_framework}
\end{figure}

We summarize the complete MTFIF-L processing flow in
Fig.~\ref{fig:overall_framework} through four stages:

\begin{enumerate}
    \item \textit{DF observation preprocessing:}
    The raw observations $\mathcal O$ are organized by time window and
    reduced to the distinct DF state sets $\widehat{\mathcal O}_k$ while
    retaining observation multiplicities.

    \item \textit{physics-guided multimode candidate generation:}
    Multimode ray tracing maps $\widehat{\mathcal O}_k$ to the
    frequency window candidate sets $\mathcal C_{f,k}$ under the
    corresponding propagation backgrounds.

    \item \textit{Time- and frequency-dependent localization:}
    window-level location extraction and same-frequency cross-window
    processing convert $\mathcal C_{f,k}$ into the frequency-specific
    competing sets $\mathcal S_f$, which are then processed by the
    frequency-dependent decision.

    \item \textit{set-valued localization output:}
    The procedure returns $\widehat{\mathcal Q}_s$; when a unique
    decision is established, the corresponding point estimate is denoted
    by $\hat{\mathbf q}_s$.
\end{enumerate}

Algorithm~\ref{alg:MTFIF-L} gives the execution sequence of the
operations defined in the preceding subsections.

\begin{algorithm}[!t]
\caption{Overall Procedure of MTFIF-L}
\label{alg:MTFIF-L}
\begin{algorithmic}[1]
\footnotesize

\REQUIRE Raw DF observations $\mathcal O$, receiver location $\mathbf q_r$,
and propagation backgrounds $\{\mathcal I_k\}_{k=1}^{K}$
\ENSURE Localization set $\widehat{\mathcal Q}_s$

\STATE Partition $\mathcal O$ into time windows
$\{\mathcal W_k\}_{k=1}^{K}$.

\FOR{$k=1,\ldots,K$}
    \STATE Apply directional refinement and repeated state aggregation
    to obtain $\widehat{\mathcal O}_k$.
    \FOR{each distinct DF state $\boldsymbol{\xi}_{u,k}
    \in\widehat{\mathcal O}_k$}
        \STATE Generate the valid $O_1$, $X_1$, $O_2$, and $X_2$
        ground landings under $\mathcal I_k$.
    \ENDFOR
    \STATE Organize the valid landings by frequency to form
    $\mathcal C_{f,k}$.
\ENDFOR

\FOR{each valid $(f,k,m)$}
    \STATE Extract
    $\mathcal H_{f,k,m}
    =\operatorname{RMin}[J_{f,k,m}]$.
\ENDFOR

\FOR{each valid $(f,k)$}
    \STATE Form
    $\mathcal H_{f,k}^{U}
    =\bigcup_{m\in\mathcal M}\mathcal H_{f,k,m}$.
\ENDFOR

\FOR{each frequency $f$ with $|\mathcal K_f|\ge1$}
    \IF{$|\mathcal K_f|\ge2$}
        \STATE Extract $\mathcal B_f^{U}$ and all valid
        $\mathcal B_{f,m}$.
        \STATE Form
        $\mathcal S_f
        =\mathcal B_f^{U}
        \cup
        \bigcup_{m:\,|\mathcal K_{f,m}|\ge2}\mathcal B_{f,m}$.
    \ELSE
        \STATE Let $k_1$ be the only valid time window and set
        $\mathcal S_f=\mathcal H_{f,k_1}^{U}$.
    \ENDIF
\ENDFOR

\STATE Construct
$\mathcal F_{\mathrm{loc}}
=\{f:\mathcal S_f\neq\varnothing\}$
and
$N_{\mathrm{loc}}=|\mathcal F_{\mathrm{loc}}|$.

\IF{$N_{\mathrm{loc}}=0$}
    \STATE $\widehat{\mathcal Q}_s\leftarrow\varnothing$.
\ELSIF{$N_{\mathrm{loc}}=1$}
    \STATE Apply the single-frequency window-wise pairwise decision
    to obtain $\widehat{\mathcal Q}_s$.
\ELSIF{$N_{\mathrm{loc}}=2$}
    \STATE Apply the two-frequency reciprocal correspondence decision
    to obtain $\widehat{\mathcal Q}_s$.
\ELSE
    \STATE Apply the multifrequency median-distance and
    frequency-wise pairwise decision to obtain
    $\widehat{\mathcal Q}_s$.
\ENDIF

\RETURN $\widehat{\mathcal Q}_s$

\end{algorithmic}
\end{algorithm}

\section{Experimental Results and Analysis}
\label{sec:results_discussion}

In this section, we evaluate MTFIF-L from three complementary perspectives: mechanism
analysis, overall localization performance, and propagation interpretation.
We first examine multimode retention and the reduction of competing
locations through same-frequency cross-window and cross-frequency
processing. We then evaluate localization accuracy and post-processing
runtime, followed by propagation-path analysis of the reported outputs.

\subsection{Experimental Setup and Evaluation Metrics}
\label{subsec:experimental_link_setting}
\label{subsec:evaluation_criteria}

\begin{figure}[htbp]
\centering
\includegraphics[width=0.78\linewidth]
{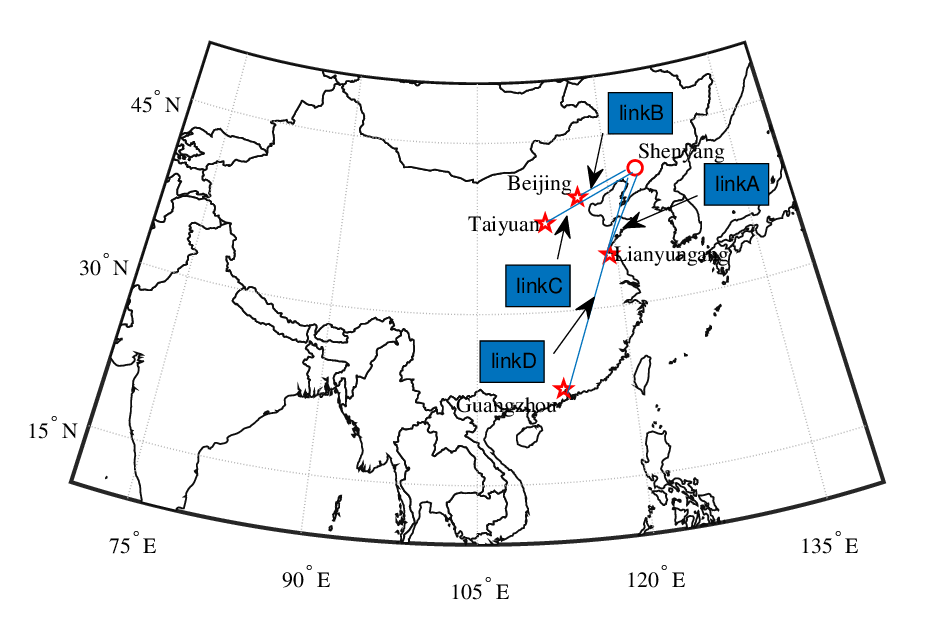}
\caption{Spatial geometry of the four measured HF links.}
\label{fig:measured_link_geometry}
\end{figure}

To evaluate MTFIF-L, we use four measured HF links sharing the same receiver
station $R$, denoted as Links A--D. The corresponding fixed emitter stations
are denoted by $T_{\mathrm A}$--$T_{\mathrm D}$. The link distances
range from $628.08$ to $2291.64$~km and include multifrequency,
two-frequency, and single-frequency observation conditions.
Fig.~\ref{fig:measured_link_geometry} shows their spatial geometry, and
Table~\ref{tab:experimental_link_setting} summarizes the principal
measurement configurations. The known emitter coordinates and bearing are
used only for post-localization evaluation and preprocessing diagnostics,
respectively, and are not involved in candidate construction or localization.

\begin{table*}[!t]
\caption{Measurement Configurations of the Four HF Links}
\label{tab:experimental_link_setting}
\centering
\begin{tabular}{c c c c c r}
\hline
Link &
Link distance (km) &
$N_f$ &
Frequency range or operating frequencies (MHz) &
Observation date &
$N_{\rm raw}$\\
\hline
A & 882.09 & 30 & 2.122--17.252 & 2025-03-03--2025-03-05 & 185465\\
B & 628.08 & 32 & 8.125--13.180 & 2025-09-25--2025-09-29 & 847696\\
C & 1022.29 & 2 & 10.000, 15.000 & 2025-03-03--2025-03-05 & 301265\\
D & 2291.64 & 1 & 17.770 & 2025-04-24--2025-04-25 & 239890\\
\hline
\end{tabular}
\begin{flushleft}
\footnotesize
Note: $N_f$ denotes the number of operating frequencies, and $N_{\rm raw}$
denotes the number of raw DF records before preprocessing. For Links A and B,
the reported frequency ranges give the minimum and maximum values among the
discrete operating frequencies and do not represent continuous frequency
sweeps.
\end{flushleft}
\end{table*}

The same preprocessing and multimode ray-tracing procedure described in
Section~\ref{subsec:physics_guided_candidate} is applied uniformly to all
four links. The DF quality field is not used in candidate construction or
localization. The propagation environment is generated using the International Reference
Ionosphere 2020 (IRI2020) and the 13th-generation International Geomagnetic
Reference Field (IGRF13), and the Haselgrove
ray equations are integrated using a fourth-order Runge–Kutta method with
the same ray-tracing configuration for all links.

The same fixed coarse-to-fine spatial search configuration is used for all
links without link-specific tuning. Its numerical settings and implementation
details are provided in the Appendix.

\paragraph{Localization Accuracy}
For a unique point estimate $\hat{\mathbf q}_s$, the absolute
localization error is defined as
\begin{equation}
e_{\rm abs}
=
d_{\rm g}
\left(
\hat{\mathbf q}_s,
\mathbf q_s
\right),
\label{eq:evaluation_abs_error}
\end{equation}
where $\mathbf q_s$ is the known emitter location and
$d_{\rm g}(\cdot,\cdot)$ denotes great-circle distance. Because the four
links have substantially different propagation distances, the relative
localization error is used as the primary accuracy metric:
\begin{equation}
e_{\rm rel}
=
\frac{
d_{\rm g}
\left(
\hat{\mathbf q}_s,
\mathbf q_s
\right)
}{
d_{\rm g}
\left(
\mathbf q_r,
\mathbf q_s
\right)
}
\times 100\%,
\label{eq:evaluation_rel_error}
\end{equation}
where $\mathbf q_r$ denotes the receiver location. The relative error
facilitates comparison across links, while $e_{\rm abs}$ reports the
corresponding spatial deviation. 
Localization errors are reported only when a unique point estimate is
obtained and are computed after localization using the known emitter
coordinates.

\paragraph{Post-Processing Runtime}
The post-processing runtime is measured after the multimode ray-tracing
candidates have been generated and therefore excludes construction of the
ionospheric propagation environment and ray tracing. It is decomposed into
local construction, within-frequency temporal processing, and final
decision:
\begin{equation}
T_{\rm total}
=
T_{\rm local}
+
T_{\rm temporal}
+
T_{\rm decision}.
\label{eq:evaluation_total_runtime}
\end{equation}

All timing tests are performed in MATLAB R2021a on a computer equipped with
an AMD Ryzen 7 8845H processor and $27.8~\mathrm{GB}$ of memory. After one
warm-up pass, each stage is executed independently ten times, and its median
runtime is reported. The total post-processing time $T_{\rm total}$ is
obtained by summing the three stage-wise medians.

\subsection{Roles of Multimode Retention, Cross-Window, and Cross-Frequency Processing}
\label{subsec:component_analysis}

We conduct three targeted experiments to examine whether physical validity
filtering eliminates propagation mode alternatives, how same-frequency
cross-window processing reduces frequency-specific competing locations,
and whether cross-frequency processing further resolves the locations that
remain. We evaluate final localization accuracy separately in the following
subsection.

\subsubsection{Role of Multimode Retention}
\label{subsec:propagation_mode_uncertainty}

Among the 226 frequency-time-window conditions of the four measured links,
225 yield physically valid landings for all four considered modes,
$O_1$, $X_1$, $O_2$, and $X_2$, corresponding to $99.56\%$ overall.
The link-specific numbers of such conditions are 95, 73, 36, and 21 for
Links A--D, respectively; these represent $98.96\%$, $100\%$, $100\%$,
and $100\%$ of the frequency-time-window conditions available for the
four links. Thus, physical validity filtering alone generally does not
eliminate the propagation mode alternatives in the measured data.

\begin{figure}[!t]
    \centering
    \includegraphics[width=\linewidth]
    {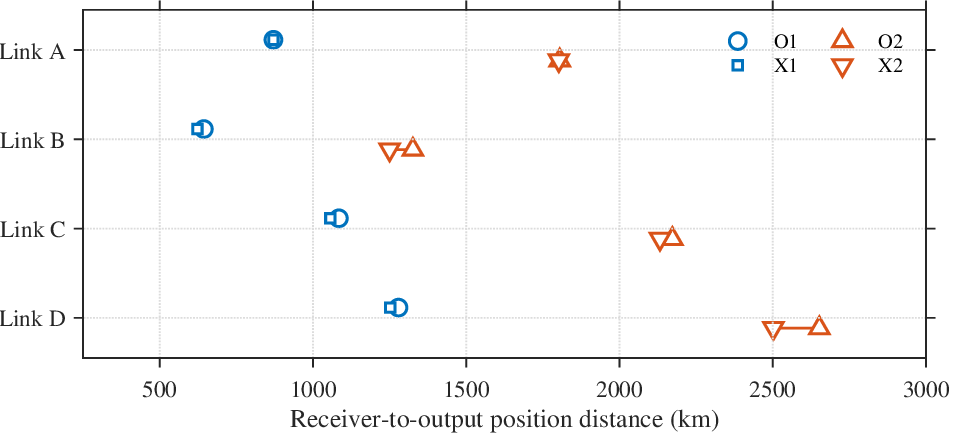}
    \caption{Receiver-to-location distance distributions for the one- and two-hop O/X propagation modes.}
    \label{fig:EXP1_candidate_distribution}
\end{figure}

The mode-separated distance distributions in
Fig.~\ref{fig:EXP1_candidate_distribution} further show that O/X modes with
the same hop count generally occupy comparable receiver-to-location ranges,
whereas one- and two-hop modes can occur at clearly different spatial scales.
For example, the O1/X1 locations of Link A are concentrated near
$871$~km, while the O2/X2 locations are near $1800$~km.

Together, these results support retaining the valid landings from all
considered propagation modes rather than preselecting a single mode before
localization.

\subsubsection{Role of Cross-Window Processing Within a Frequency}
\label{subsec:temporal_observation_role}

To isolate the role of cross-window processing, each operating frequency is
processed independently using all of its valid time windows, without
introducing information from other frequencies. Across the four links,
65 frequency-specific conditions are available. Of these, 52 contain at
least two valid time windows, and 20 yield a unique single-frequency
decision after cross-window processing, as summarized in
Table~\ref{tab:single_frequency_temporal_decision}.

\begin{table}[!t]
\centering
\caption{Cross-Window Disambiguation Under Per-Frequency Isolation}
\label{tab:single_frequency_temporal_decision}
\footnotesize
\setlength{\tabcolsep}{3.5pt}
\renewcommand{\arraystretch}{1.08}
\begin{tabular}{c c c c}
\hline
Link &
\shortstack{No. of\\frequencies} &
\shortstack{With $\geq 2$ valid\\time windows} &
\shortstack{Unique single-frequency\\decisions} \\
\hline
A & 30 & 25 & 8  \\
B & 32 & 24 & 10 \\
C & 2  & 2  & 1  \\
D & 1  & 1  & 1  \\
\hline
Total & 65 & 52 & 20 \\
\hline
\end{tabular}
\end{table}

Link C provides a useful counterexample to a simple window count
interpretation: 10 MHz remains unresolved with 19 valid time windows,
whereas 15 MHz yields a unique result with 17 windows. Decision uniqueness
therefore depends on the geographic information contained in the
window-level location sets rather than on the number of time windows alone.

The frequency-specific sets that remain unresolved after cross-window
processing are retained for the subsequent frequency-dependent decision.
The localization accuracy of the frequency-specific decisions is examined
separately in Section~\ref{subsubsec:per_frequency_accuracy}.

\subsubsection{Role of Cross-Frequency Processing}
\label{subsec:cross_frequency_disambiguation}

To evaluate the role of cross-frequency processing, the frequency-specific
competing location sets of Links A--C are processed jointly across their
available operating frequencies.
Under per frequency isolation, unique locations are obtained for 8, 10,
and 1 operating frequencies in Links A, B, and C, respectively,
corresponding to $26.67\%$, $31.25\%$, and $50.00\%$ of the
frequencies available for the three links. Nevertheless, the subsequent
cross-frequency decision produces a unique final location for all three links.
Thus, cross-frequency processing does not require every individual frequency
to be uniquely resolved beforehand; unresolved frequency-specific sets can
still contribute to the final decision.

\subsection{Localization Performance and Post-Processing Efficiency}
\label{subsec:localization_performance}

This subsection first reports the final localization performance on the
four measured links, followed by a fixed-mode direct ray-tracing reference
and a further accuracy evaluation of the frequency-specific decisions.
The post-processing runtime is then examined.

\subsubsection{Measured Link Localization Performance}
\label{subsec:measured_link_localization_performance}

Table~\ref{tab:localization_results} summarizes the final decision,
receiver-referenced location, and post-localization error for each link.

\begin{table}[!t]
\centering
\caption{Localization Results for the Four Measured Links}
\label{tab:localization_results}
\scriptsize
\setlength{\tabcolsep}{2.3pt}
\renewcommand{\arraystretch}{1.05}
\begin{tabular}{c c c c c c c}
\toprule
Link &
$N_f$ &
Status &
\shortstack{Range\\(km)} &
\shortstack{Bearing\\($^\circ$)} &
\shortstack{$e_{\rm abs}$\\(km)} &
\shortstack{$e_{\rm rel}$\\(\%)} \\
\midrule
A & 30 & Unique & 872.00  & 212.22 & 12.32   & 1.40  \\
B & 32 & Unique & 633.00  & 259.50 & 4.92    & 0.78  \\
C & 2  & Unique & 1056.49 & 263.32 & 34.90   & 3.41  \\
D & 1  & Unique & 1252.00 & 216.28 & 1041.32 & 45.44 \\
\bottomrule
\end{tabular}

\vspace{1mm}
\parbox{\columnwidth}{\scriptsize
Note: $N_f$ denotes the number of frequencies used for decision.
Range and bearing are referenced to the receiver. Localization errors are
computed using the known emitter coordinates only after the location
decision and are not involved in localization.}
\end{table}

All four links satisfy their corresponding uniqueness conditions and
therefore yield unique location decisions. Links A--C achieve relative
localization errors between $0.78\%$ and $3.41\%$, whereas Link D exhibits
a substantially larger relative error of $45.44\%$.

Link D contains only one operating frequency. Its 21 valid time windows
produce 42 competing locations after same-frequency cross-window processing,
from which the full-window strict pairwise rule selects a unique location.
The resulting relative error is $45.44\%$. Since Link D is the only
single-frequency measured link, this result alone is insufficient to assess
MTFIF-L under single-frequency observation conditions. This issue is
examined further through the per-frequency results of Links A--C below.

\subsubsection{Fixed-Mode Direct Ray-Tracing Reference}
\label{subsubsec:fixed_mode_direct_baseline}
To characterize the sensitivity of direct location inversion to the
prescribed propagation mode, four fixed-mode references are considered
using $O_1$, $X_1$, $O_2$, and $X_2$, respectively. For each observation,
a valid ground landing under the prescribed mode is directly taken as the
source location estimate, and the mean relative localization error is
computed over all valid observations with observation multiplicities
restored. These results are used as a reference for prescribed-mode
inversion rather than as a like-for-like estimator benchmark for MTFIF-L.

The fixed-mode results show a clear dependence on
both the prescribed propagation mode and link distance.
Links A--C have link distances of $882.09$, $628.08$, and $1022.29$~km,
respectively; for these links, the one-hop modes yield substantially
smaller errors than the two-hop modes. In contrast, Link D has a
substantially longer link distance of $2291.64$~km, for which the two-hop
modes perform better. This indicates that no single prescribed mode is
consistently suitable across different links. Moreover, although the valid
landing rates for Links A--C exceed $98\%$ for all four modes, their
localization errors differ markedly, showing that a valid ray landing alone
does not identify the propagation mode appropriate for localization.

\begin{table}[!t]
\centering
\caption{Mean Relative Error (\%) of Fixed-Mode Ray-Tracing Localization}
\label{tab:fixed_mode_direct_baseline}
\footnotesize
\setlength{\tabcolsep}{6.0pt}
\renewcommand{\arraystretch}{1.05}
\begin{tabular}{c c c c c}
\toprule
Link & $O_1$ & $X_1$ & $O_2$ & $X_2$ \\
\midrule
A & 23.42 & 24.01 & 88.71  & 79.76  \\
B & 30.26 & 27.31 & 137.28 & 125.57 \\
C & 9.31  & 9.09  & 96.94  & 93.97  \\
D & 67.30 & 59.29 & 48.62  & 44.54  \\
\bottomrule
\end{tabular}
\end{table}

These results demonstrate the strong dependence of direct inversion on the
prescribed propagation mode. No single mode is consistently favorable
across the four links, and physical validity alone does not identify the
mode that provides the smaller localization error. The fixed-mode results
therefore serve primarily to illustrate the limitation of prescribing one
propagation mode before localization.

\subsubsection{Localization Accuracy Under Per-Frequency Isolation}
\label{subsubsec:per_frequency_accuracy}

To assess MTFIF-L under single-frequency conditions, we evaluate the
localization accuracy of the frequency-specific decisions for Links A--C.
Among the 51 frequencies with at least two valid time windows, 19 yield
unique decisions, corresponding to $37.3\%$; these include 8, 10, and 1
frequencies for Links A, B, and C, respectively.

\begin{table}[!t]
\centering
\caption{Localization Accuracy Under Per-Frequency Isolation}
\label{tab:per_frequency_accuracy}
\scriptsize
\setlength{\tabcolsep}{3.0pt}
\renewcommand{\arraystretch}{1.05}
\begin{tabular}{c c c c c}
\toprule
Link &
\shortstack{Unique /\\Eligible} &
\shortstack{Median\\(\%)} &
\shortstack{Range\\(\%)} &
\shortstack{Full MTFIF-L\\(\%)} \\
\midrule
A & 8/25  & 4.58 & 1.60--114.44 & 1.40 \\
B & 10/24 & 9.97 & 0.63--121.96 & 0.78 \\
C & 1/2   & 4.67 & 4.67         & 3.41 \\
\bottomrule
\end{tabular}

\vspace{1mm}
\parbox{\columnwidth}{\scriptsize
Note: Eligible frequencies contain at least two valid time windows.
Median and range are calculated over the unique frequency-specific
decisions only.}
\end{table}

Among the unique frequency-specific decisions, the median relative errors
are $4.58\%$, $9.97\%$, and $4.67\%$ for Links A--C, respectively, while
the corresponding ranges show substantial variation across individual
frequencies. Thus, temporal information alone can provide unique and
low-error localization under some single-frequency conditions, although
both resolving capability and accuracy depend on the information available
at the individual frequency. In comparison, the full MTFIF-L results for
Links A--C are $1.40\%$, $0.78\%$, and $3.41\%$, indicating that
cross-frequency information provides additional robustness against
unreliable frequency-specific outcomes.

\subsubsection{Post-Processing Computational Efficiency}
\label{subsubsec:computational_efficiency}
Table~\ref{tab:computational_efficiency} summarizes the candidate processing
scales and stage-wise post-processing runtimes obtained using the timing
protocol defined in Section~\ref{subsec:evaluation_criteria}. 

Links B and D contain the largest ray candidate sets, with $370816$ and
$142044$ candidates, respectively, and local construction is also the
dominant stage for these two links. It accounts for approximately $89.95\%$
and $89.80\%$ of their total post-processing times. For Links A and C,
within-frequency temporal processing is the largest stage, accounting for
approximately $48.38\%$ and $57.14\%$ of the total runtime, respectively.

The final frequency-dependent decision is not the dominant computational
stage for any of the four links. Its largest contribution occurs for Link A,
where it accounts for approximately $11.36\%$ of the total post-processing
time, while the remaining links show smaller fractions. The observed runtime
differences therefore arise primarily before the final decision, during local
construction and within-frequency temporal processing.

\begin{table}[!t]
\centering
\caption{MTFIF-L Candidate Scale and Post-Processing Runtime}
\label{tab:computational_efficiency}
\scriptsize
\setlength{\tabcolsep}{2.0pt}
\renewcommand{\arraystretch}{1.05}
\begin{tabular}{c r r r r r r r}
\toprule
Link &
$N_{\rm ray}$ &
$N_{\rm win}$ &
$N_{\rm freq}$ &
$T_{\rm local}$ &
$T_{\rm temp}$ &
$T_{\rm dec}$ &
$T_{\rm total}$ \\
\midrule
A & 24996  & 1027 & 478 & 14.40  & 17.30 & 4.062 & 35.77  \\
B & 370816 & 557  & 440 & 185.81 & 14.59 & 6.169 & 206.57 \\
C & 6280   & 305  & 43  & 3.87   & 5.47  & 0.233 & 9.58   \\
D & 142044 & 315  & 42  & 93.03  & 10.54 & 0.028 & 103.60 \\
\bottomrule
\end{tabular}

\vspace{1mm}
\parbox{\columnwidth}{\scriptsize
$N_{\rm ray}$: ray candidates;
$N_{\rm win}$: window-level locations;
$N_{\rm freq}$: frequency-specific competing locations.
All runtimes are in seconds.}
\end{table}

\subsection{Propagation-Based Interpretation of the Localization Results}
\label{subsec:propagation_physics_interpretation}

This subsection provides a post-localization interpretation of the
ray-tracing paths associated with the reported locations, focusing on their
propagation mode composition and its variation across operating frequencies.

\subsubsection{Dominant Propagation Paths and Physical Characteristics}
\label{subsec:propagation_mode_physics}

For Links A--C, the ray-tracing paths associated with the reported locations
exhibit distinct propagation compositions. Link A contains one-hop E,
one-hop F, and two-hop FF paths, whereas Link B is dominated by one-hop F
propagation, which accounts for $83.91\%$ when the O/X branches are combined.
For Link C, only one-hop X mode paths are retained, with F-layer and E-layer
propagation accounting for $65.21\%$ and $34.79\%$, respectively. Although
the E- and F-layer groups of Link C occupy different reflection height and
elevation angle ranges, both are associated with the same reported location.

For Link D, the reported location has a large spatial error, and its
propagation interpretation is therefore evaluated separately using the true
link geometry. In this truth-referenced analysis, one-hop O/X F-layer paths
account for approximately $46.50\%$, while two-hop O/X FF paths account for
approximately $43.45\%$. These results show that the true link geometry
admits several physically feasible one-hop/two-hop and O/X propagation
interpretations. The true emitter location is used only for this
post-localization propagation analysis and is not involved in candidate
construction or the MTFIF-L localization decision. Accordingly, this
analysis should not be interpreted as propagation support for the selected
$1252$~km localization result.

\subsubsection{Frequency-Dependent Propagation Characteristics}
\label{subsec:frequency_path_composition}

\begin{figure*}[!t]
    \centering
    \includegraphics[width=0.98\textwidth]
    {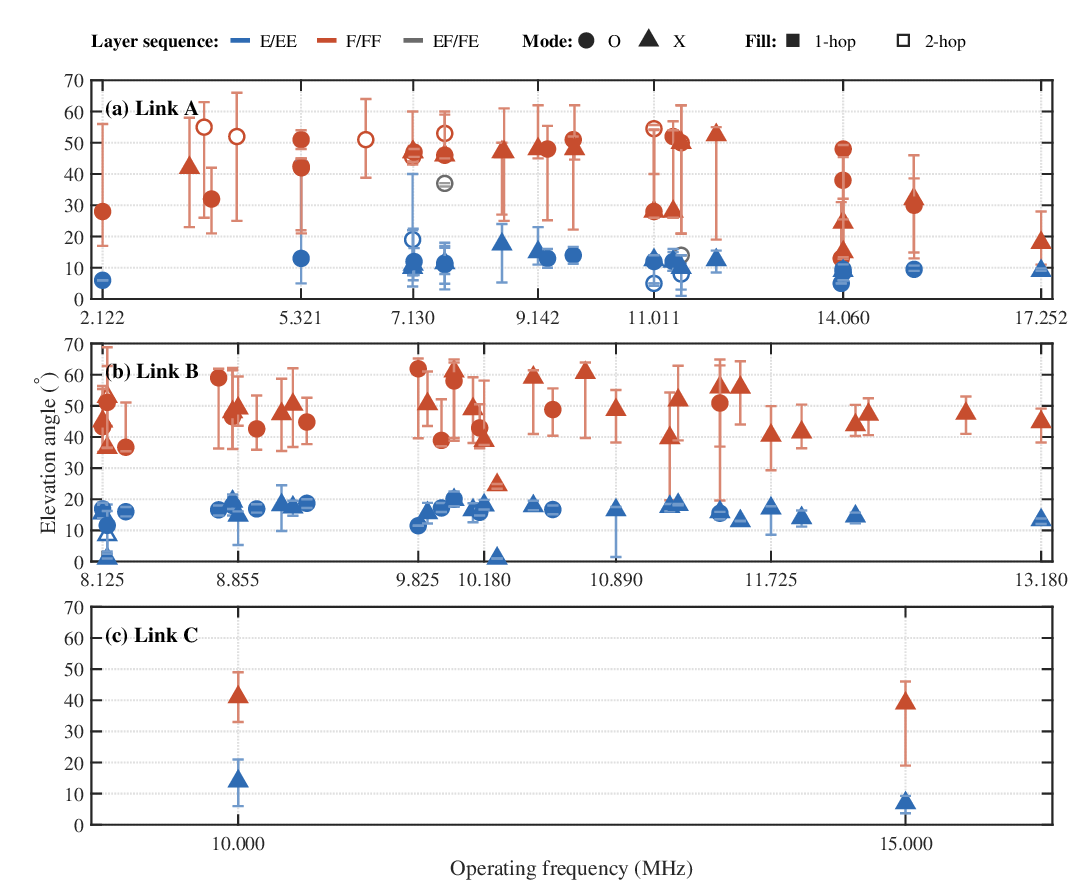}
    \caption{Propagation-path types and elevation-angle distributions
    across operating frequencies for Links A--C. Markers indicate the median
    ($P_{50}$), with vertical bars spanning the 10th--90th percentiles
    ($P_{10}$--$P_{90}$).}
    \label{fig:frequency_elevation_mode_v6}
\end{figure*}

Figure~\ref{fig:frequency_elevation_mode_v6} shows the propagation path 
types and elevation angle distributions across the operating frequencies 
of the multifrequency Links A--C. The reflection layer sequence, 
O/X magnetoionic mode, and hop count all vary with operating frequency. 
Propagation paths whose ionospheric hops are confined to the F region 
generally occur at higher elevation angles than those confined to 
the E-region, whereas mixed-region two-hop paths involving 
one E-region hop and one F-region hop are observed only at
selected frequencies.

The frequency dependence is not deterministic. Propagation paths involving
E-layer reflections and those involving F-layer reflections occur over
overlapping frequency ranges, indicating that the operating frequency alone
does not uniquely determine the reflection-layer composition of the path. 
For Link C, the corresponding frequency-resolved composition
statistics show that the F-layer fraction increases from $57.7\%$ at
$10$~MHz to $86.0\%$ at $15$~MHz. Link B remains predominantly associated
with one-hop propagation, whereas several lower-frequency conditions of
Link A contain two-hop paths. The O/X composition also varies with
frequency, without a common monotonic trend across the measured links.

These results show that different operating frequencies can probe the same
stationary emitter through different propagation realizations. This provides
a physical basis for treating the frequency-specific candidate sets as
complementary evidence in MTFIF-L, rather than as repeated observations of
a single fixed propagation path.

\subsection{Discussion}
\label{subsec:discussion}

The experiments support the successive use of multimode retention,
temporal processing, and cross-frequency processing in MTFIF-L. Physical
validity filtering alone generally does not remove the propagation-mode
alternatives, whereas cross-window processing reduces the competing
locations within each frequency. For multifrequency links, the remaining
frequency-specific sets can be further combined even when individual
frequencies are not uniquely resolved. The observed frequency-dependent
propagation characteristics further indicate that different frequencies
provide complementary rather than redundant propagation information.

The per-frequency isolation results show that temporal information alone
can support unique and low-error localization under some single-frequency
conditions, although both decision availability and localization accuracy
vary across frequencies. Thus, the relatively large error of Link D should
not be interpreted as a general failure of single-frequency MTFIF-L, while
a unique decision should not itself be regarded as a measure of absolute
accuracy. When multiple frequencies are available, cross-frequency
processing provides additional evidence and reduces dependence on
potentially unreliable frequency-specific outcomes.

The evaluation is limited to four measured HF links, and the additional
single-frequency cases are obtained by isolating frequencies from Links
A--C rather than from independent single-frequency links. Moreover, the
measured data do not provide propagation-mode labels or additional
component-resolved observables for direct comparison with methods requiring
such information, and the reported runtime excludes ionospheric environment
construction and ray tracing. Future work will extend the evaluation to
additional measured links and investigate improved propagation information
and calibrated localization uncertainty.

\section{Conclusion}
\label{sec:conclusion}

We developed MTFIF-L for HF single-station localization when the
propagation mode composition is unavailable from conventional DOA
measurements. We retain physically feasible multimode ray candidates and
use same-frequency cross-window processing and, when available,
cross-frequency processing to determine the transmitter location without
requiring the propagation mode composition to be identified a priori.

Experiments on four measured HF links yielded unique location decisions
for all four links. Links A--C achieved relative localization errors of
$1.40\%$, $0.78\%$, and $3.41\%$, respectively, whereas Link D exhibited
a substantially larger error of $45.44\%$. The per-frequency results show
that temporal information can support localization under single-frequency
conditions, while cross-frequency information provides additional
robustness against unreliable frequency-specific outcomes. Propagation
analysis further reveals link-dependent variations in reflection layer,
hop count, magnetoionic mode, and elevation angle, providing a physical
interpretation of the localization results.

Future work will evaluate the method using measured links covering
additional geographical regions, propagation distances, and ionospheric
conditions, and will investigate the extent to which more accurate
ionospheric backgrounds and additional independent propagation observations
can further improve localization accuracy.

\appendix[Numerical Details of the MTFIF-L Localization Core]
\label{supp:mtfif_numerical}

This appendix specifies the numerical search and decision rules used by
MTFIF-L. The common spatial-search configuration is given in
Table~\ref{tab:MTFIF-L_numerical_configuration}.

\begin{table}[!t]
\centering
\footnotesize
\setlength{\tabcolsep}{3.2pt}
\renewcommand{\arraystretch}{1.10}
\caption{Numerical Search Configuration of MTFIF-L}
\label{tab:MTFIF-L_numerical_configuration}
\begin{tabular}{
p{0.45\columnwidth}
p{0.28\columnwidth}}
\toprule
Configuration & Value \\
\midrule
Coarse-search range step
& $20~\mathrm{km}$
\\
Coarse-search bearing step
& $0.5^\circ$
\\
Fine range step $\Delta r_{\rm fine}$
& $1~\mathrm{km}$
\\
Fine bearing step $\Delta\beta_{\rm fine}$
& $0.05^\circ$
\\
Search-domain expansion
& $1$ coarse-grid cell
\\
Local-refinement half-width
& $\pm40~\mathrm{km}$, $\pm1^\circ$
\\
Maximum recentering iterations
& $4$
\\
\bottomrule
\end{tabular}
\end{table}

\subsection{Regional-Minimum Extraction}
For each objective $J(\mathbf q)$, $\operatorname{RMin}[J]$ is evaluated
over a receiver-centered polar search domain constructed from the spatial
extent of the locations supplied to the current stage. The range interval is
expanded according to Table~\ref{tab:MTFIF-L_numerical_configuration}, and
the bearing interval is formed from the circular span to avoid the
$0^\circ/360^\circ$ discontinuity.

On the coarse grid, let $J_{ij}=J(\mathbf q(r_i,\theta_j))$ and let
$\mathcal N_8(i,j)$ denote the valid eight-neighbor set of node $(i,j)$.
For a finite set of numerical values $\mathcal X$, define
\begin{equation}
\tau(\mathcal X)
=
64\,\operatorname{eps}
\left(
\max\left\{1,\max_{x\in\mathcal X}|x|\right\}
\right),
\label{eq:numerical_tolerance}
\end{equation}
where $\operatorname{eps}(x)$ denotes double-precision floating-point
spacing near $x$; $\tau(\mathcal X)$ is used only for numerical equivalence
and is not a physical association threshold. Let $\mathcal X_J$ contain all
finite coarse-grid values of $J$. A finite node $(i,j)$ is a
regional-minimum candidate when
\begin{equation}
J_{ij}
\leq
J_{uv}+\tau(\mathcal X_J),
\qquad
\forall (u,v)\in\mathcal N_8(i,j).
\label{eq:regional_minimum_numerical_rule}
\end{equation}
If none is detected while finite grid values exist, the finite global minimum
is used as a fallback seed.

For $J_{f,k,m}$ and $J_f^{U}$, each connected equal-valued plateau is
represented by one coarse seed, whereas the regional-minimum nodes of
$J_{f,m}$ and $J_{\rm CF}$ are refined separately. The former avoids
duplicate representations of the same flat minimum region, while the latter
avoids premature merging before local refinement. Each seed is refined on
the local fine grid in Table~\ref{tab:MTFIF-L_numerical_configuration}; if
the minimum lies on the grid boundary, the grid is recentered until an
interior minimum is obtained or the maximum iteration count is reached.
Refined locations falling in the same fine-grid range--bearing cell are
collapsed by retaining the member with the smallest objective value. The
surviving locations constitute $\operatorname{RMin}[J]$. Coincident refined
locations are similarly collapsed after combining $\mathcal B_f^{U}$ and
the valid $\mathcal B_{f,m}$ sets.

\subsection{Numerical Handling in the Final Decision}
All pairwise decisions use strict distance comparisons. In the
single-frequency branch, exact equality alone produces an abstention and no
numerical distance margin is introduced. In the two-frequency reciprocal
search, the subsequent two-frequency time-window comparisons, and the
pairwise comparisons for three or more frequencies, floating-point
equivalence is identified using $\tau(\mathcal X)$ in
\eqref{eq:numerical_tolerance}; distances within this tolerance are treated
as ties.

For the two-frequency branch, reciprocal-pair midpoints are computed on the
unit sphere rather than by directly averaging latitude and longitude. Let
$\mathbf u_1$ and $\mathbf u_2$ be the corresponding three-dimensional unit
vectors. Their spherical midpoint is
\begin{equation}
\mathbf u_{\mathrm m}
=
\frac{\mathbf u_1+\mathbf u_2}
{\left\|\mathbf u_1+\mathbf u_2\right\|_2},
\end{equation}
and $\mathbf u_{\mathrm m}$ is converted back to latitude and longitude.

\bibliographystyle{IEEEtran}
\bibliography{references}
\end{document}